\documentstyle[12pt]{article}
\begin{document}

\begin{titlepage}
\begin{flushright}
\hfil                 UCSD/PTH 92-39\\
\hfil November 1992 \\
\end{flushright}
\begin{center}

\vspace{1cm}
{\LARGE\bf Chiral Fermions, Anomalies and Chern-Simons Currents on the
Lattice \footnotemark[1]}

\vspace{2.0cm}
{\large Karl Jansen}\\
\vspace{1cm}
University of California at San Diego\\
Department of Physics-0319 \\
La Jolla, CA~92093-0319 \\
USA  \\
\vspace{0.2cm}

\vspace{2.0cm}
\end{center}

\abstract{
\noindent
I discuss the zeromode spectrum of lattice chiral fermions
in the domain wall model suggested recently. In particular I give the
critical momenta where the fermions cease to be chiral and show
that the spectrum is directly related to the behaviour of the
Chern-Simons current on the lattice. First results for the domain wall
model on the finite lattice indicate that the relevant features of the
model in the infinite system survive for the finite lattice.
}
\vfill
\footnotetext[1]{{\em Talk presented at the International Conference on
Lattice Field Theory, LATTICE 92, (1992) Amsterdam}}
\end{titlepage}

\section{Introduction}
In this talk I want to report some first results for the domain wall
model suggested by Kaplan \cite{david}.
The basic idea of this model is to start with an odd dimensional --and
therefore vectorlike-- theory and to
add a mass term for the fermions which depends on the
extra dimension and has the form of a soliton or kink,
generating in this way a
domain wall. It is well known that in a situation like this
one finds zeromodes bound to the domain wall \cite{jarebbi}. Kaplan was
able to show that these zeromodes represent chiral fermions on the lattice and
that the unwanted doubler modes can be removed by introducing the usual
Wilson-term. These results \cite{david,martin}, which hold for the free theory
and for the infinite lattice, could be demonstrated to survive even on
a finite lattice \cite{karl}. In addition the cancellation of
the zeromode anomaly by the divergence of the Goldstone-Wilczek current
as described in \cite{david} could also been seen
on the finite lattice \cite{karl}.

In this talk I will discuss the peculiar behaviour of the
zeromode spectrum. I will give the critical momenta of these zeromodes
where they lack to be chiral and relate the zeromode spectrum to the
behaviour of the Chern-Simons current on the lattice.

\section{The infinite lattice}

To be specific I will discuss the spectrum and also the anomaly
for the case of a
3-dimensional model. The results generalized to
arbitrary dimensions
can be found in \cite{martin,maarten}. I start with the Dirac-Wilson operator
on an infinite lattice with lattice spacing $a=1$

\begin{equation}
K_{3D}  =\sum_{\mu=1}^3\sigma_\mu\partial_\mu
+m\epsilon(s)+\frac{r}{2}\sum_{\mu=1}^3\Delta_\mu
\label{eq:diracco}
\end{equation}
\noindent where $\partial$ denotes the lattice derivative
$\partial_\mu = \frac{1}{2}\left[\delta_{z,z+\mu} - \delta_{z,z-\mu}\right] $,
$\Delta$ the lattice Laplacian
$\Delta_\mu = \left[\delta_{z,z+\mu} + \delta_{z,z-\mu}-2\delta_{z,z}\right] $,
the $\sigma_\mu$ are the usual Pauli matrices and
$r$ the Wilson coupling.
I will denote by $s$ the extra dimension along which the mass defect
appears, while $x,t$ are the 2-dimensional coordinates.
The domain wall is taken to be a step function $\epsilon$,
\begin{equation}
\epsilon(s) = \left\{ \begin{array}{lll}
                    -1 & s < 0 \\
                     0 & s=0\\
                    +1 & s > 0
                     \end{array} \right.
\label{eq:epsilon}
\end{equation}
\noindent where the height of the domain wall is given by the mass parameter
$m$ which I will choose to be positive throughout this paper.

We are looking for solutions which are plane waves in the $(x,t)$-plane
\begin{equation}
\Psi^{\pm} = e^{i(p_tt+p_xx)}\Phi(s)u^{\pm}
\label{eq:psi}
\end{equation}
where $u^\pm$ are the eigenspinors of $\sigma_3$,
$\sigma_3 u^\pm = \pm u^\pm$.
With this ansatz the Dirac operator becomes
\begin{equation}
K_{3D} = \sum_{i=1}^2i\sigma_i\sin(p_i)+ \sigma_3\partial_s
+m\epsilon(s)
       + r\sum_{i=1}^2(\cos(p_i)-1)+\frac{r}{2}\Delta_s  \;\; .
\label{eqnarray:diracmom}
\end{equation}
The final goal is to diagonalize the 3 dimensional Dirac operator
in such a way that it reduces to the 2 dimensional Dirac operator
for free massless fermions,$K_{3D}\Psi = K_{2D}\Psi$
where $K_{2D}$ acting on $\Psi$ is given by
\begin{equation}
K_{2D} = \sum_{\mu=1}^2\sigma_\mu\partial_\mu = i(\sigma_1 \sin(p_t)
+\sigma_2 \sin(p_x))\;.
\label{eq:e1}
\end{equation}
\noindent Hence the equation to solve is
\begin{equation}
\left[\sigma_3\partial_s + m\epsilon(s) -rF
+\frac{r}{2}\Delta_s\right]\Phi u^\pm =0
\label{eq:problem}
\end{equation}
where $F=\sum_{i=t,x}(1-\cos(p_i))$.
Following \cite{david} I choose an exponential ansatz for $\Phi$
away from the domain wall $\Phi(s+1)=z\Phi(s)$.
Inserting this into (\ref{eq:problem}) one finds four solutions
\begin{equation}
z =\frac{r-m_{eff}\pm\sqrt{m_{eff}(m_{eff}-2r)+1}}{r\pm1}
\label{eq:roots}
\end{equation}
where $m_{eff} = m\epsilon(s)-rF$, the $\pm$
in the nominator stand for the two roots and the $\pm$ in the
denominator stand for the chirality.
Note that in the limit $r=1$ eq.(\ref{eq:roots}) can be reduced to the
corresponding expressions in \cite{david}.

One has to impose the condition that the solutions are normalizable to
obtain sensible wavefunctions. This means
that
$|z|>1$ for $s<0$ and $|z|<1$ for $s>0$.
The boundaries of the regions where chiral solutions exist are obtained
by setting $|z|=1$.
Explicit matching of the normalizable solutions for
positive and negative $s$ at $s=0$
enables one to determine the regions with chiral fermions.
One finds that existence and chirality of the solutions is independent of
the sign of $r$
and that a negative $m$ leads to opposite chiralities.
Depending on the values of $m/r$ one gets
$m =rF$ and $m =r(F+2)$ as the boundaries for the critical momenta,
where $F$ is defined as above.

The results can be summarized as follows.
Starting with $m/r=0$ one finds no chiral fermions. For increasing $0<m/r<2$
the region in momentum space around $\vec{p}=(0,0)$ where chiral modes exist
grows. This region is bounded by $m=rF$ which gives the upper critical
momenta.
Increasing $m/r$ above $m/r=2$ opens up the two ``doubler''
modes at $\vec{p} =(0,\pi)$ and $\vec{p} =(\pi,0)$ which have flipped
chirality, while the original mode at $\vec{p} =(0,0)$ disappears.
Here the boundaries of the regions in momentum space are given by $m=rF$ for
the
lower and $m=r(F+2)$ for the upper critical momenta.
For $m/r>4$ the two ``doublers'' disappear and one gets a zero mode at
$\vec{p} =(\pi,\pi)$ with the same chirality as the mode
at $\vec{p} =(0,0)$.
The boundary for the lower critical momenta is
given by $m=r(F+2)$. This mode is finally
also lost as $m/r$ is increased to $m/r \ge 6$.

It should be remarked that this spectrum stems from $\Psi^+$
solutions only, and that
there are no $\Psi^-$ solutions for positive $m$. The
change of the chirality is the usual reinterpretation of the
chirality at different corners of the Brillouin zone.

The zero mode spectrum as found here is directly related to the
coefficient of the lattice Chern-Simons current
induced by heavy fermions. This current is
responsible for the anomaly cancellation. As there is a chiral zeromode
bound to the domain wall, there exists the corresponding anomaly.
However, we started with a vectorlike theory and consequently
the theory should be anomaly free.
As is explained in \cite{david} this contradiction is resolved by the
fact that the domain wall induces a Goldstone-Wilczek current \cite{gw}
far off the domain wall the divergence of which exactly cancels the
zeromode anomaly on the domain wall \cite{caha}.

Recently, this current was also calculated on
the lattice.
The radiatively induced Chern-Simons action in d=3 dimensions is given
by \cite{redlich}
\begin{equation}
\Gamma_{CS}=\epsilon_{\mu\nu\rho}\int d^3x A_\mu\partial_\nu A_\rho\; .
\label{eq:csaction}
\end{equation}
The effective action is then given by $c\Gamma_{CS}$. The coefficient
$c$ can be calculated in perturbation theory
\cite{costeluescher,maarten} and is given by
\begin{equation}
c = i\epsilon_{\mu\nu\rho}\frac{\partial}{\partial (q)_\nu} \left.
I\right|_{q=0}
\label{eq:c1}
\end{equation}
with $I$ the following integral
\begin{equation}
I = \int \frac{d^3p}{(2\pi)^3}Tr\left[ S(p)\Lambda_\mu(p,p-q)\right.
    \left. S(p-q) \Lambda_\rho(p+q,p)\right]
\label{eqnarray:int1}
\end{equation}
where $S$ is the fermion propagator and $\Lambda_\mu$ denotes the
photon vertex.
Imposing the Ward identity
\begin{equation}
\Lambda_\mu(p,p)= -i\partial/\partial
p_\mu S^{-1}(p)
\end{equation}
the relevant integral becomes
\begin{equation}
\int\frac{d^3p}{(2\pi)^3}Tr
                      \left[S\partial_\mu S^{-1}\right]
                      \left[S\partial_\nu S^{-1}\right]
                      \left[S\partial_\rho S^{-1}\right]\; .
\label{eq:c2}
\end{equation}

The fermion propagator can be written in a generic form,
$S^{-1}(p)=a(p)+i\vec{b}(p)\vec{\sigma}$ which has the structure
of $S^{-1}(p) = N(p)V(p)$, where $V(p)$ is a $SU(2)$ matrix and $N(p)$ a
normalization factor. Provided that $S^{-1}$ does not vanish for any $p$
the integral eq.(\ref{eq:c2}) does not depend on $N(p)$ and
has a simple topological interpretation: It is the
winding number of the map $V$ from the torus $T^3$ to $SU(2)$ or the sphere
$S^3$. (In general this map is a map from the torus $T^d$ to the sphere
$S^d$.)

Taking the usual Wilson fermion propagator the winding number can be calculated
by using the fact that the integral has to be computed only
in infinitesimal regions near the Brillouin corners.
One obtains (see \cite{maarten} for more details)
\begin{equation}
c= \frac{-i}{32\pi^3}\sum_{k=0}^3 (-1)^k\frac{m-2rk}{\left| m-2rk\right|}\;
{}.
\label{eq:c3}
\end{equation}
This expression should be compared with the (unregulated) continuum result
\cite{caha}
\begin{equation}
c_{cont}= \frac{-i}{32\pi^3}\frac{m}{|m|}\; .
\label{eq:ccont}
\end{equation}
Therefore one finds that for $m/r<0$, $c=0$. For $m/r<2$ the lattice
result is twice the continuum value. For $2<m/r<4$ it is $-4c_{cont}$ and
for $4<m/r<6$ it is again twice the continuum value.

For the case of the domain wall model in which we are interested here
the result has the following implication:
For the calculation of the Goldstone-Wilczek current one goes far off
the domain wall and assumes there the mass to be constant. Then one
can calculate the current with the method described above. For $m/r<2$
one finds therefore that the current only flows on one side of the
domain wall and has twice the value of the continuum result. Of course,
the divergencies will come out the same.
The behaviour of the lattice Chern-Simons coefficient
finds its exact correspondence in the zeromode spectrum as
discussed above where we found 1 lefthanded, 2 righthanded and 1
lefthanded chiral fermion for $0<m/r<2$, $2<m/r<4$ and $4<m/r<6$,
respectively.

\section{And the Finite Lattice}

Any numerical work on this system will necessarily involve finite lattices,
and so I now compare the results obtained on the infinite
system with the ones of a finite lattice.
On the finite lattice one has to choose
some boundary conditions. Taking periodic boundary conditions
generates a second anti-domain wall.
The mass term is therefore modified to be $m\epsilon_L(s)$ with
\begin{equation}
\epsilon_L(s) = \left\{ \begin{array}{lll}
                    -1 & 2\le s \le \frac{L_s}{2} \\
                    +1 & \frac{L_s}{2}+2 \le s \le L_s \\
                     0 & s=1, \frac{L_s}{2}+1
                     \end{array} \right. \;\; .
\label{eq:m0}
\end{equation}

The zeromodes on the finite lattice can be searched for by solving the
Hamiltonian problem numerically.
If one again assumes plane waves in the $x$-direction the
Hamiltonian is given by \cite{karl}
\begin{equation}
H  = -\sigma_1\left[i\sigma_2\sin(p_x) + \sigma_3\partial_s
+m\epsilon_L(s)\right.
+ \left. r(\cos(p_x)-1) +\frac{r}{2}\Delta_s) \right] \;\; .
\label{eqnarray:hamiltonian}
\end{equation}

The eigenvalues and eigenfunctions of the Hamiltonian
eq.(\ref{eqnarray:hamiltonian}) were calculated numerically~.
To find the critical momenta the ratio
\begin{equation}
R=\frac{{\bar\Psi}\Psi}{{\bar\Psi}\sigma_1\Psi}
\label{eq:R}
\end{equation}
was studied,
which is a normalized measure for whether the fermions are chiral or not. It is
zero
if the fermions are chiral and $R>0$ for non-chiral modes (see
fig.2b in \cite{karl}). To determine whether one still has chiral
fermions a threshold value for R was defined. If $R < 0.01$
the fermions were regarded to be chiral.

Comparing the results from the infinite system
with the finite lattice calculations with $L=100$ \cite{martin},
one finds that the two are
practically indistinguishable.
For $L=20$, a lattice size realistic for simulations, a
small shift occurs. Fixing $m$ and $r$ we find for $m/r<2$ a smaller value
and for $2<m/r<4$ a larger value of the critical momentum.

It is also possible to extract the Chern-Simons current on the finite
lattice in the presence of a smooth external gauge field configuration.
The current is most easily computed by the inverse fermion matrix of
the model which can be obtained by standard methods like conjugate
gradient.
Note that this is an exact solution of the numerical problem
and that no simulation is involved \cite{karl}.
%\epsfxsize=5.0 cm
%\centerline{\epsffile{fig1.ps}}
%\centerline{{\bf Figure 1:} Figure 1}

I show in fig.~1 the Chern-Simons current on a $16^3$ lattice with
$m=0.81$ and $r=0.9$. Note that these values of $m,r$ are quite large.
The picture nicely demonstrates that the advocated flow of the current
as obtained above is reproduced on the finite lattice: It flows only on
one side of the domain wall and is zero on the other.

It is a simple task to get the divergence of the current and one can
demonstrate \cite{karl} that it obeys the continuum anomaly equation to
a very good accuracy. It is also possible to  find non-trivial anomaly
cancellation like in the 3-4-5 model where the individual fermion
currents are anomalous but the sum of them cancel.

In summary the domain wall model shows a lot of promising features on
the finite lattice like the chiral zeromode spectrum and the correct
anomaly behaviour. Although these properties were only found for the
free theory or with weak external gauge fields they certainly point into the
right direction and are encouraging. The next step is to include
dynamical gauge fields and see whether one can find the desired
behaviour.
\section*{Acknowledgements}
I want to thank M.F.L. Golterman, D. Kaplan and M. Schmaltz
for giving me the opportunity
to present our work at this conference. I also want to thank them and
J. Kuti for numerous helpful and stimulating discussions.
This work is supported by DOE grant DE-FG-03-90ER40546.

\pagebreak

\section*{Figure Caption}

{\bf Fig.1}
I show the Chern-Simons current in arbitrary units
as obtained on a $16^3$
lattice. It shows the expected peculiar behaviour that it flows only on
one side of the domain wall and is zero on the other. The locations of
the domain walls is at s=1 and at s=9.

\end{document}